\documentclass[twocolumn,aps,prb,floatfix,superscriptaddress]{revtex4-2}
\usepackage{amsmath}
\usepackage{amsfonts}
\usepackage{amssymb}
\usepackage{graphicx}
\usepackage{color}
\usepackage{bm}

\usepackage{soul}

\usepackage{mathtools}

\DeclarePairedDelimiterX\braket[2]{\langle}{\rangle}{#1 \delimsize\vert #2}
\usepackage[final]{hyperref} % adds hyper links inside the generated pdf file
\hypersetup{
	colorlinks=true,       % false: boxed links; true: colored links
	linkcolor=blue,        % color of internal links
	citecolor=blue,        % color of links to bibliography
	filecolor=magenta,     % color of file links
	urlcolor=blue         
}
\newcommand{\be}{\begin{equation}}
\newcommand{\ee}{\end{equation}}
\newcommand{\bea}{\begin{eqnarray}}
\newcommand{\eea}{\end{eqnarray}} 
\newcommand{\mc}{\mathcal}
\newcommand{\mb}{\mathbf}

\newcommand{\new}[1]{\textcolor{black}{#1}}

\begin{document}

\title{Tunneling spectroscopy of the spinon-Kondo effect in one-dimensional Mott insulators} 

\author{Rodrigo G. Pereira}
\affiliation{International Institute of Physics and Departamento de F{\'i}sica Te{\'o}rica e Experimental, Universidade Federal do Rio Grande do Norte, Natal, Rio Grande do Norte, 59078-970, Brazil}

\author{Bruno F. Marquez}
\affiliation{Centro At{\'o}mico Bariloche, CNEA, Bariloche, Argentina}
\author{Karen Hallberg}
\affiliation{Centro At{\'o}mico Bariloche, CNEA, Bariloche, Argentina}
\affiliation{Instituto Balseiro and Instituto de Nanociencia y Nanotecnolog{\'i}a, CNEA-CONICET, Bariloche, Argentina}

\author{Tim Bauer}
\affiliation{Helmholtz-Zentrum Berlin für Materialien und Energie, Hahn-Meitner-Platz 1, 14109 Berlin, Germany}
\affiliation{Dahlem Center for Complex Quantum Systems and Institut f\"ur Theoretische Physik,
Freie Universit\"at Berlin, Arnimallee 14, 14195 Berlin, Germany}

\author{Reinhold Egger}
\affiliation{Institut f\"ur Theoretische Physik, Heinrich-Heine-Universit\"at, D-40225  D\"usseldorf, Germany}

\begin{abstract}
We study the tunneling density of states (TDOS) in  one-dimensional (1D) Mott insulators at energies below the charge gap.  By employing nonlinear Luttinger liquid theory and density-matrix renormalization group (DMRG) simulations, we predict that in the presence of a magnetic impurity at the boundary, characteristic Fermi-edge singularity features can appear at subgap energies in the TDOS near the boundary.  In contrast to the Kondo effect in a metal, these resonances are strongly asymmetric and of power-law form. The power-law exponent is universal and determined by the spinon-Kondo effect.  
\end{abstract}
\maketitle

\section{Introduction} \label{sec1}

It is well known that a spin-$1/2$ magnetic impurity in a metal causes a Kondo resonance pinned to the Fermi level, with the ground-state impurity spin fully screened by conduction electrons \cite{Hewson1993}.  The Kondo effect implies a robust zero-energy peak in the TDOS near the impurity that
can be observed in scanning tunneling spectroscopy (STS) or in transport experiments. 
We  consider here the TDOS near a magnetic impurity in a \emph{Mott insulator}, which has no spin gap but a finite charge gap $\Delta$ due to Umklapp processes for interacting lattice fermions with commensurate band filling. Since the state after a tunneling event must accomodate an extra electron or hole, which entails
both the spin and charge degrees of freedom, a conventional zero-energy Kondo peak is ruled out by the charge gap. Can a magnetic impurity nonetheless induce subgap (energy $|E|<\Delta$) resonance features in the TDOS of a Mott insulator?  
This question is of key importance for recent STS experiments conducted on 2D spin liquid candidates, depositing spin-$1/2$ 
Co atoms on 1T-TaSe$_2$ \cite{Chen2022} or spin-$3/2$ molecules on 1T-NbSe$_2$ \cite{Zhang2024}.
Theoretical work \cite{Law2017,He2018}  suggests that these materials may host a quantum spin liquid phase characterized by a spinon Fermi surface, with experimental evidence supporting this scenario \cite{Ribak2017,Gomilsek2019,Murayama2020,Chen2025}. In such  gapless quantum spin liquids,  parton mean-field theories  \cite{Khaliullin1995,Kolezhuk2006,Florens2006,Ribeiro2011,Zheng2024} predict a spinon-Kondo effect, where spinons---rather than electrons---screen  magnetic impurities. The observation of subgap  features by STS  \cite{Chen2022,Zhang2024} has been attributed to a combination of this spinon-Kondo effect and  interactions between spinons and gapped charge degrees of freedom (``holons'' or ``doublons''), which together bind these fractional excitations into a localized electron-like mode \cite{He2022}. Importantly, the spinon-Kondo effect should also occur in 1D Mott insulators. Experimentally available candidate platforms include nanographene chains \cite{Mishra2021,Zhao2024a,Zhao2025} (see also \cite{Jacob2021}), molecular chains \cite{Sun2025,Su2025}, and van der Waals materials \cite{Park2024}. In 1D systems, spin-charge separation naturally emerges at low energy scales \cite{Gogolin1998}, and the existence of a Kondo screening cloud formed by spinons around a boundary impurity spin is well established in spin-$1/2$ Heisenberg  chains  \cite{Eggert1992,Wang1997,Furusaki1998,Laflorencie2008,Kattel2024,Zhakenov2025,Moca2025,Kulka2025}.  However, to address how spinons affect the TDOS,  
one must include the charge degrees of freedom.

\begin{figure}
\begin{center}
   \includegraphics[width=0.8\columnwidth]{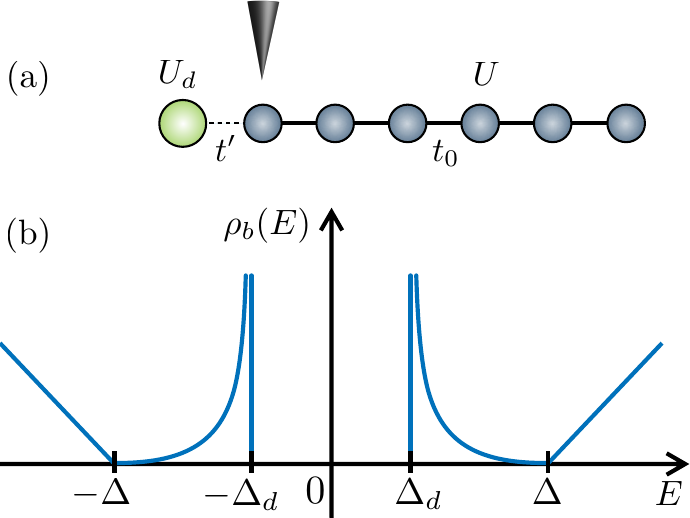}
    \caption{ Schematic setup and TDOS near a boundary magnetic impurity in a 1D Mott insulator. (a) As example, we consider a half-filled Hubbard chain with on-site interaction strength $U>0$ and tunnel coupling $t_0$, connected by the coupling $t'$ to a boundary Anderson impurity with interaction $U_d>0$, see Eq.~\eqref{model}.  We compute the TDOS $\rho_\sigma(j,E)$ in Eq.~\eqref{tdos}, which can be measured by STS via a probe tip, shown here at site $j=1$. (b) Boundary TDOS $\rho_{\rm b}(E)$ taken at $j=1$, where $\Delta$ $\,(\Delta_d$) is the bulk (impurity) charge gap. The subgap resonance features for $\Delta_d<|E|<\Delta$  are power-law Fermi-edge singularities, see Eq.~\eqref{subgapregime}, where the power-law exponent $\alpha=1/2$ is caused by the spinon-Kondo effect.
    \new{We here assume that $\Delta_d$ and $\Delta$ are well separated such that the subgap TDOS contribution becomes very small for $|E|\approx \Delta$.  In that case, the linear TDOS energy dependence for $|E|>\Delta$ near the boundary is also  observable in the presence of the magnetic impurity.} }    \label{fig1}
\end{center}
\end{figure}

In this paper, we study the energy-dependent subgap TDOS near a magnetic impurity in a 1D Mott insulator, see Fig.~\ref{fig1}(a). 
We mainly focus on the case where the impurity is located at the boundary. Results for an impurity embedded in the bulk are given in the Appendix.
The 1D case allows for nonperturbative analytical results by means of nonlinear Luttinger liquid theory \cite{Imambekov2012,Schmidt2010,Pereira2012,Essler2015}, which we compare to numerical DMRG simulation \cite{White1992} results.  In this manner, we obtain essentially exact results for the TDOS, where the uncontrolled approximations  
introduced by parton mean-field theories in higher dimensions are avoided altogether.  Our analytical results are based on the universal low-energy theory of 1D Mott insulators,
valid for arbitrary commensurate band fillings and/or microscopic lattice realizations. 
Numerical DMRG results are obtained for a half-filled particle-hole symmetric Hubbard chain (sites $j=1,\ldots, L$) connected to a boundary Anderson impurity (site $j=0$), see Fig.~\ref{fig1}(a),
\bea
H&=&-t_0\sum_\sigma\sum_{j=1}^{L-1} (c^\dagger_{j\sigma}c^{\phantom\dagger}_{j+1,\sigma}+\text{h.c.})\nonumber\\
&+&U\sum_{j=1}^L\left(n_{j\uparrow}-\frac12\right)\left(n_{j\downarrow}-\frac12\right) \label{model} \\ \nonumber
&-&t'\sum_\sigma(c^\dagger_{0\sigma}c^{\phantom\dagger}_{1\sigma}+\text{h.c.})+U_d \left(n_{0\uparrow}-\frac12\right)\left(n_{0\downarrow}-\frac12\right),
\eea
with $n_{j\sigma}=c^\dagger_{j\sigma}c_{j\sigma}^{}$ and the electron annihilation operator $c_{j\sigma}$ at site $j$ with spin projection $\sigma$. We note that for arbitrary $U/t_0$, the charge gap of an infinitely long Hubbard chain is given by \cite{Essler2005,Pereira2012}
\be\label{gap}
\Delta=\frac{8t_0^2}{U}\int_1^{\infty}dw\, \frac{\sqrt{w^2-1}}{\sinh(2\pi w t_0/U)}.
\ee
For $(t_0,t')\ll (U,U_d)$, a boundary magnetic impurity is then coupled to a 1D Mott insulator, where $\Delta\sim U$ from Eq.~\eqref{gap}.
In general, on energy scales below $\Delta$, the physics is captured by an effective spin model $H_{\rm spin}$. For the Hubbard example, one obtains \cite{Kattel2024,Kattel2025n} 
\be\label{spinchain}
H_{\rm spin}=J_{\rm imp}\mb S_0\cdot \mb S_1+J\sum_{j\ge 1}\mb S_j\cdot \mb S_{j+1},
\ee
where $\mb S_j$ are spin-$1/2$ operators and the exchange couplings are  $J=4t_0^2/U$ and $J_{\rm imp}=8t^{\prime\, 2}/(U+U_d)$.

Including the relevant charge modes not contained in $H_{\rm spin}$, we focus on the parameter regime $U_d\alt 2\Delta$ and study the site- and energy-dependent TDOS, 
\be \label{tdos}
\rho_\sigma(j,E)=\int_{-\infty}^{\infty}\frac{dt}{2\pi}\, e^{i E t}\langle\{c^{\phantom\dagger}_{j\sigma}(t),c^\dagger_{j\sigma}(0)\}\rangle,
\ee
with the anticommutator $\{\cdot,\cdot\}$.  For particle-hole symmetric models, 
$\rho_\sigma(j,-E)=\rho_\sigma(j,E)$. Moreover, spin SU(2) invariance implies  
$\rho_\uparrow(j,E)=\rho_\downarrow(j,E)$. In particular, we define the impurity  TDOS, $\rho_{\rm i}(E)=\rho_\sigma(j=0,E)$, and the  boundary TDOS, $\rho_{\rm b}(E)=\rho_\sigma(j=1,E)$.  

Our key results are summarized in Fig.~\ref{fig1}(b), which shows the schematic energy dependence of the boundary TDOS. For above-gap energies, we obtain a linear increase, $\rho_{\rm b}(E) \sim (|E|-\Delta)\Theta(|E|-\Delta)$, with the Heaviside step function $\Theta$.  Without impurity, the boundary TDOS vanishes for $|E|<\Delta$. 
For arbitrary 1D Mott insulators, we predict characteristic subgap resonances in the presence of the impurity. Instead of a Kondo peak, we find strongly asymmetric Fermi-edge singularity peaks for $\Delta_d<|E|<\Delta$, see Fig.~\ref{fig1}(b),
\begin{equation}\label{subgapregime}
    \rho_{\rm i,b}(E)\sim (|E|-\Delta_d)^{-\alpha}\Theta(|E|-\Delta_d).
\end{equation} 
A rough estimate gives the threshold energy $\Delta_d\approx U_d/2$, but
more detailed estimates are provided below. 
The singularity exhibits a universal exponent $\alpha=1/2$ within \new{an energy window $|E|-\Delta_d \ll T_K, \Delta-\Delta_d$}, where $T_K$ is the Kondo temperature  from the spinon-Kondo effect. (For the Hubbard case,  $T_K\sim e^{-\frac{\pi}2\sqrt{J_{\rm imp}/J}}$  for $J_{\rm imp}\ll J$ \cite{Laflorencie2008,Kattel2024}.) The value $\alpha=1/2$ follows directly from SU(2) spin symmetry, which fixes the scaling dimension of the operator associated with spinons. The subgap resonance features \eqref{subgapregime} are observed in our DMRG simulations, and we find that the exponent extracted from numerical results for finite chains approaches the analytical prediction as we increase the coupling to the impurity. 

The remainder of this paper is structured as follows. In Sec.~\ref{sec2}, we present an effective field theory based on a nonlinear Luttinger liquid theory, and show how the scenario sketched in Fig.~\ref{fig1} can be understood from analytical arguments.  The strong-coupling limit is addressed in Sec.~\ref{sec3}.  In Sec.~\ref{sec4}, we turn to DMRG simulation results and compare them to the analytical results.  Finally, we discuss our findings in Sec.~\ref{sec5}. The Appendix contains technical details and additional explanations.

\section{Effective field theory}\label{sec2} 

We first present analytical arguments based
on a nonlinear Luttinger liquid theory that  describes threshold singularities in 1D Mott insulators \cite{Pereira2012,Essler2015}.  
We start with  $t'=0$. Within bosonization, $c_{j\sigma}$ is expressed in terms of left- and right-moving charge and spin ($\nu=c,s$) boson fields
$\varphi_{\nu,L/R}(x)$  \cite{Gogolin1998}.  Here we take the continuum limit, $ja\to x$, with lattice spacing $a$.
We impose open boundary conditions by folding bosons, $\varphi_{c,L}(x)=\varphi_{c,R}(-x)+\sqrt{\pi}$ and $\varphi_{s,L}(x)=\varphi_{s,R}(-x)$, such that one
can work with chiral (say, right-moving) bosons, $\varphi_\nu(x)\equiv \varphi_{\nu,R}(x)$, 
defined on the full line $-\infty<x<\infty$. They obey the algebra $[\varphi_{\nu}(x),\varphi_{\nu'}(x')]=\frac{i}2 {\delta_{\nu\nu'}\text{sgn}(x-x')}$.  

Using the chiral currents $J_{\nu}^z=-\frac{1}{\sqrt{4\pi}}\partial_x\varphi_{\nu}$ and  
$J_{\nu}^{\pm}=\frac{1}{2\pi}e^{\pm i\sqrt{4\pi}\varphi_{\nu}}$, we obtain the spin-charge-separated low-energy Hamiltonian   \cite{Gogolin1998,Essler2005}
\begin{equation}
H=\sum_{\nu}\int_{-\infty}^\infty dx \left[\frac{2\pi v_\nu}3 \mb J_{\nu}^2   -2\pi \lambda_\nu v_\nu \mb J_{\nu}(x) \cdot  {\mb J}_{\nu}(-x)  \right]. 
\end{equation}
For a half-filled Hubbard chain with $|U|\ll t_0$, one finds $v_{c,s}=2t_0\pm \frac{U}{2\pi}$ and $\lambda_{c,s}=\mp \frac{U}{2\pi t_0}$ \cite{Essler2002}. For repulsive interactions, the spin coupling $\lambda_s$ is marginally irrelevant, whereas $\lambda_c$ is marginally relevant and causes a charge gap consistent with Eq.~\eqref{gap}.
The  picture of gapped charge and gapless spin sectors remains qualitatively correct as the interaction strength increases. For strong interactions,  the   spin sector is described by $H_{\rm spin}$ with spinons as elementary excitations, while charge excitations (holons)  correspond to gapped spinless fermions \cite{Penc1995,Matveev2007}.

Following Ref.~\cite{Pereira2012}, we describe holons as mobile charge impurities in 
high-energy bands, restricting the Fock space to contain at most one holon. 
Nonlinear Luttinger liquid theory \cite{Pereira2012,Essler2015} then expresses the (say, spin-$\uparrow$) electron operator as $c^\dagger_{j,\uparrow} \sim  d^\dagger_c(x)e^{i\sqrt{\pi}\varphi_{s}(x)}$, where
 $d^\dagger_c$ creates a holon in the upper Hubbard band and the vertex operator $e^{i\sqrt{\pi}\varphi_s}$ creates  a low-energy spinon.  In contrast with fermionic spinons in parton mean-field theories \cite{He2022}, here the vertex operator associated with a chiral spin-1/2 excitation has scaling dimension $1/4$ and obeys semionic statistics \cite{Pham2000}.
 The dynamics is now captured by an effective mobile holon impurity model,
\be\label{MIM}
H_{\rm MIM}=\frac{v_s}{2}\int_{-\infty}^{\infty} dx(\partial_x\varphi_{s})^2+\int_0^{\infty}dx\, d^\dagger_c\left(\Delta-\frac{\partial_x^2}{2m_c}\right)d_c^{\phantom\dagger},
\ee
where  $m_c$ is the holon mass and we impose  $d_c(0)=0$. 
Time reversal symmetry, which takes $\varphi_s\mapsto-\varphi_s$,   implies that the leading bulk spin-charge coupling term is $\sim d^\dagger_c d^{\phantom\dagger}_c(\partial_x\varphi_s)^2$. However, this irrelevant coupling neither generates contributions for $|E|<\Delta$ nor modifies the leading  energy dependence of the TDOS for $E\agt \Delta$. 
One can thus compute the TDOS  directly using Eq.~\eqref{MIM}, where $\rho_{\rm b}(E)$ follows by setting $d_c(a)\sim a\partial_x d_c(0)$. 
For $t'=0$, we   obtain $\rho_{\rm b}(E)\sim (|E|-\Delta)\,{\Theta(|E|-\Delta)},$ see Fig.~\ref{fig1}(b).  
A related calculation for the bulk of the chain ($j\gg 1$) instead gives
a step edge, $\rho_\sigma(j,E)\sim \Theta(|E|-\Delta)$, consistent with Ref.~\cite{Essler2002}. 

Next, we add the impurity coupling $t'\ne 0$, see Fig.~\ref{fig1}(a), and study the subgap regime $|E|<\Delta$.  
We here assume $\Delta_d\ll \Delta$ but comparison with DMRG results discussed below indicates that this assumption can be relaxed.
Performing a Schrieffer-Wolff transformation to lowest nontrivial order in $t'/\Delta$ while allowing for charge excitations 
at the magnetic impurity site only, we arrive at the transformed low-energy Hamiltonian
\bea\label{HKlow}
&& \tilde H_{\rm eff}= U_d \left(\frac12-2(S_0^z)^2\right)+\frac{v_s}{2}\int_{-\infty}^{\infty}dx\, (\partial_x\varphi_s)^2\\\nonumber &&-2\pi v_s\lambda_{\parallel}S_0^z\partial_x\varphi_s(0)+
\pi v_s\lambda_\perp \left(c^{\dagger}_{0\uparrow}c^{\phantom\dagger}_{0\downarrow}e^{-i\sqrt{4\pi}\varphi_s(0)}+\text{h.c.}\right),
\eea
with $S_0^z=(n_{0\uparrow}-n_{0\downarrow})/2$ and dimensionless 
Kondo couplings $\lambda_{\parallel,\perp}$. The sector with an empty or doubly occupied impurity site ($S_0^z=0$) is separated from the sector with a singly occupied site ($S_0^z=\pm 1/2$) by the energy difference $U_d/2$. 
Moreover, the transformed boundary electron operator
is $\tilde c_{1\sigma}= c_{1\sigma}+ a_0 c_{0\sigma}$. Boundary correlators such as $\langle \tilde c^{}_{1\sigma}(t) \tilde c_{1\sigma}^\dagger(0)\rangle$ therefore pick up the contribution $a_0^2 \langle c^{\phantom\dagger}_{0\sigma}(t)c^\dagger_{0\sigma}(0)\rangle$ from the impurity site.
For the spin-isotropic Hubbard case, we have $\lambda_\parallel=\lambda_\perp=\frac{J_{\rm imp}}{\pi v_s}$ and $a_0=\frac{2t'}{U+U_d}$. 

Momentarily neglecting the $\lambda_\perp$-term in Eq.~\eqref{HKlow}, 
we arrive at the standard Fermi-edge singularity problem \cite{Gogolin1998} by applying the unitary transformation
$W=\exp\left(- i \frac{4\gamma}{\sqrt\pi} S_0^z  \varphi_s(0)\right)$, 
which acts as 
\begin{equation}
W^\dagger c_{0\sigma}W=c_{0\sigma} \, e^{-i\sigma\sqrt{4\pi}(\gamma/\pi) \varphi_s(0)},
\end{equation}
and eliminates the $\lambda_\parallel$-term from Eq.~\eqref{HKlow} by choosing
$\gamma=\frac{\pi^{3/2}}{2}\lambda_\parallel$. While $\lambda_\parallel\propto J_{\rm imp}$ at weak coupling, 
more generally $\gamma$ corresponds to the exact scattering phase shift off the boundary since $W$ is a boundary-condition-changing operator \cite{Affleck1994}. It is then straightforward to show   
\be
\rho_{\rm i}(E)\sim \frac{\Theta(|E|-\Delta_d)}{(|E|-\Delta_d)^{1-2(\gamma/\pi)^2}}. \label{plaw}
\ee
At the same time, the boundary TDOS develops a subgap contribution,  $\rho_{\rm b}(E)\sim  a_0^2 \rho_{\rm i}(E)$.  

Let us now include the $\lambda_\perp$ term in Eq.~(\ref{HKlow}). The unitary transformation $W$ acts nontrivially on this term,
\be
W^\dagger c^{\dagger}_{0\uparrow}c^{\phantom\dagger}_{0\downarrow}e^{-i\sqrt{4\pi}\varphi_s(0)} W=c^{\dagger}_{0\uparrow}c^{\phantom\dagger}_{0\downarrow}e^{-i\sqrt{4\pi}(1-2\gamma/\pi)\varphi_s(0)}.
\ee
Since $\gamma>0$, the vertex operator becomes more relevant. 
In fact, within a perturbative RG  approach restricted to the $(S_0^z)^2=1/4$ sector, 
the Kondo couplings flow \cite{Gogolin1998} toward 
the strong-coupling fixed point $\lambda_{\perp}=\lambda_{\parallel}\to \infty$ with
phase shift $\gamma=\pi/2$. 

To verify this picture, we employ an argument similar to the Friedel sum rule in the bosonization approach to the X-ray edge problem \cite{Gogolin1998}. 
First, at the boundary, the spin-raising operator corresponds to $J_s^+(0)\sim e^{i\sqrt{4\pi}\varphi_s(0)}$ and the magnetization is $S^z_{\rm b}=-\frac1{\sqrt{4\pi}}\int_{-\epsilon}^{\epsilon}dx\, \partial_x\varphi_s(x)$,
where $\epsilon$ is a short-distance scale and the prefactor follows from $[S^z_{\rm b},J_s^{+}(0)]=J_s^{+}(0)$.  Using the relation 
\be
W^\dagger \partial_x\varphi_s(x)W=\partial_x\varphi_s(x)+\frac{4\gamma}{\sqrt{\pi}}S_0^z \delta(x),
\ee
we   obtain 
\be 
\langle S^z_{\rm b}\rangle=-\frac{2\gamma}{\pi}\langle S_0^z\rangle.
\ee
For $\gamma=\pi/2$,
we arrive at a strong-coupling Kondo singlet, $\langle S^z_{\rm b}+S^z_0 \rangle=0$.
The exchange interaction in Eq.~(\ref{HKlow}) thus leads to the spinon-Kondo effect.
Note that the transformed vertex operator then reduces to the identity. As a consequence,
Eq.~\eqref{plaw}  gives the correct result for the impurity  TDOS in the strong-coupling limit.  Inserting $\gamma=\pi/2$ into Eq.~\eqref{plaw}, we arrive at Eq.~\eqref{subgapregime} which holds close to the threshold, $0<|E|-\Delta_d\ll T_K$.   

We note that higher-order terms in $t'/\Delta$ generate impurity contributions to the TDOS at sites $j>1$. The subgap TDOS weight is then expected to decay exponentially with the distance from the boundary on a length scale $\sim v_s/\Delta$.

\section{Strong-coupling limit}\label{sec3}

\begin{figure}[t]
\includegraphics[width=0.8\columnwidth]{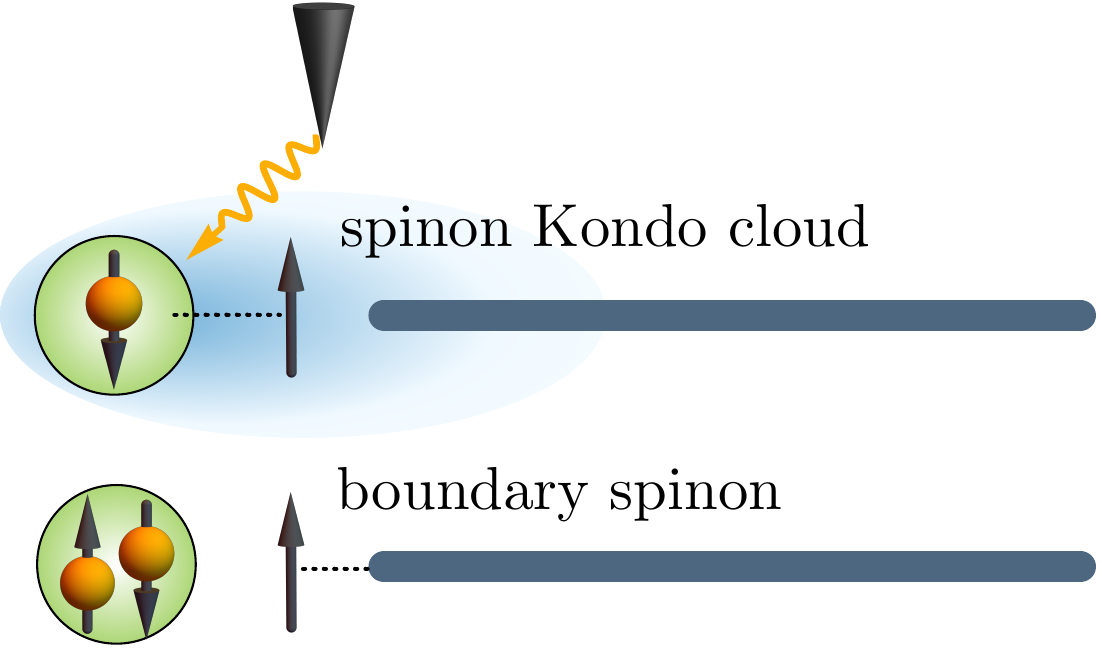}
\caption{At the strong-coupling fixed point (top), a spinon from the bulk screens the magnetic moment of  the electron at the impurity site $j=0$. 
   After applying $c_{0\uparrow}^\dagger$,  the two electrons at $j=0$ form a singlet (bottom).  As a consequence, the Kondo  singlet is broken and a spinon is added to the chain boundary. } \label{fig2} 
\end{figure}

The square-root TDOS singularity \eqref{subgapregime} can be rationalized by the strong-coupling picture illustrated in Fig.~\ref{fig2}. 
In addition to the energy cost $U_d/2$ for adding an electron to the impurity site, the tunneling process entails breaking the Kondo singlet which costs an energy $\delta E_s\sim T_K$. 
Furthermore, charge fluctuations due to $t'\neq 0$ also renormalize $\Delta_d$ by an energy $\delta E_c$, resulting in 
\be
\Delta_d = U_d/2 +\delta E_s+\delta E_c.
\ee
We estimate $\delta E_c$ by describing holons through a tight-binding model of noninteracting spinless fermions \cite{Penc1995,Matveev2007}, where one expects a holon
bound state localized near the boundary with energy $U_d/2+\delta E_c$ and $\delta E_c<0$. For small $t'$, we find $|\delta E_c|\gg \delta E_s$
such that $\Delta_d<U/2$, see the Appendix for details.  However, for larger $t'$, perturbative expressions break down and $\delta E_s>|\delta E_c|$ becomes possible, resulting in $\Delta_d>U_d/2$. 

 In any case, near the strong-coupling fixed point, the  boundary electron fractionalizes according to 
 \be
 c_{1\uparrow}^\dagger \sim d^\dagger_{\rm bs} e^{i\sqrt\pi\varphi_s(0)},
 \ee
 where $d^\dagger_{\rm bs}$ occupies the holon bound state and the vertex operator injects a spinon back into the chain, see Fig.~\ref{fig2}.  Since the vertex operator has scaling dimension 1/4, its time correlations decay $\propto 1/\sqrt t$, and a Fourier transform gives the $1/\sqrt{|E|-\Delta_d}$ singularity in Eq.~\eqref{subgapregime}.  Importantly, $e^{\pm i\sqrt\pi\varphi_s}$ is a nonlocal operator, which is why it does not appear in previous discussions of the Kondo effect in spin chains \cite{Laflorencie2008}. However, it enters in the single-electron spectral function in combination with the operator $d^\dagger_{\rm bs}$ acting in the charge sector.   In a sense, the TDOS therefore probes the ``local spinon spectral function'' of a Mott insulator. 

\section{DMRG results}\label{sec4}

\begin{figure}[t]
        \includegraphics[width=.85\columnwidth]{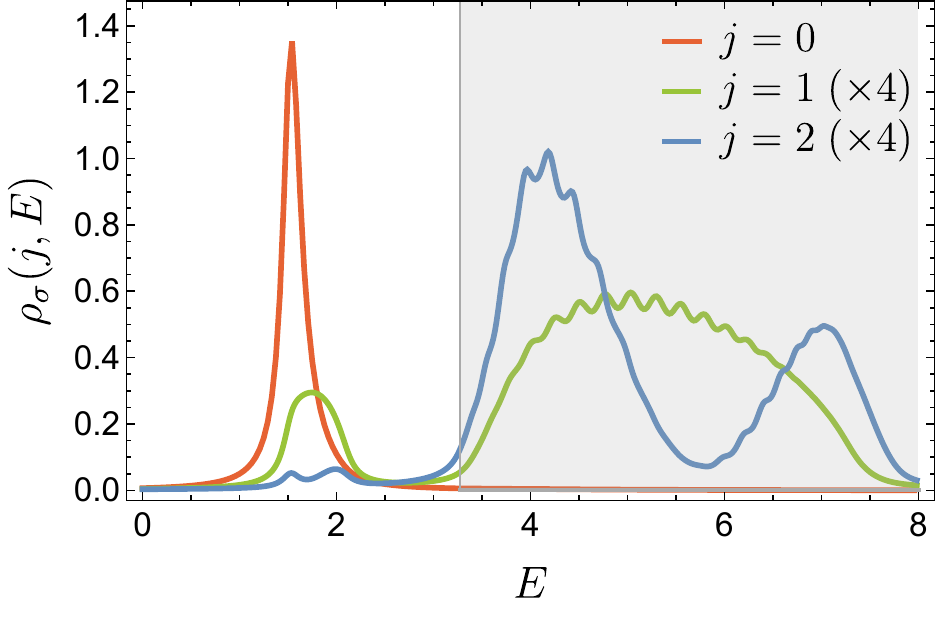}
    \caption{DMRG results for the energy dependence of the TDOS at sites $j=0,1,2$ for the model \eqref{model}, using 
      $L=23$ with $t_0=1$, $U=10$, $U_d=3$, and $t'=0.6$. The Lorentzian broadening is $\eta=0.1$. The shaded region marks the continuum of bulk states with $E>\Delta$. \new{We note that for each site $j$, the TDOS integrated over all energies gives the same value; note that here the curves for $j=1,2$ have been rescaled. The corresponding sum rule is discussed in the Appendix.} } \label{fig3}
\end{figure} 

\begin{figure}[t]
\includegraphics[width=.99\columnwidth]{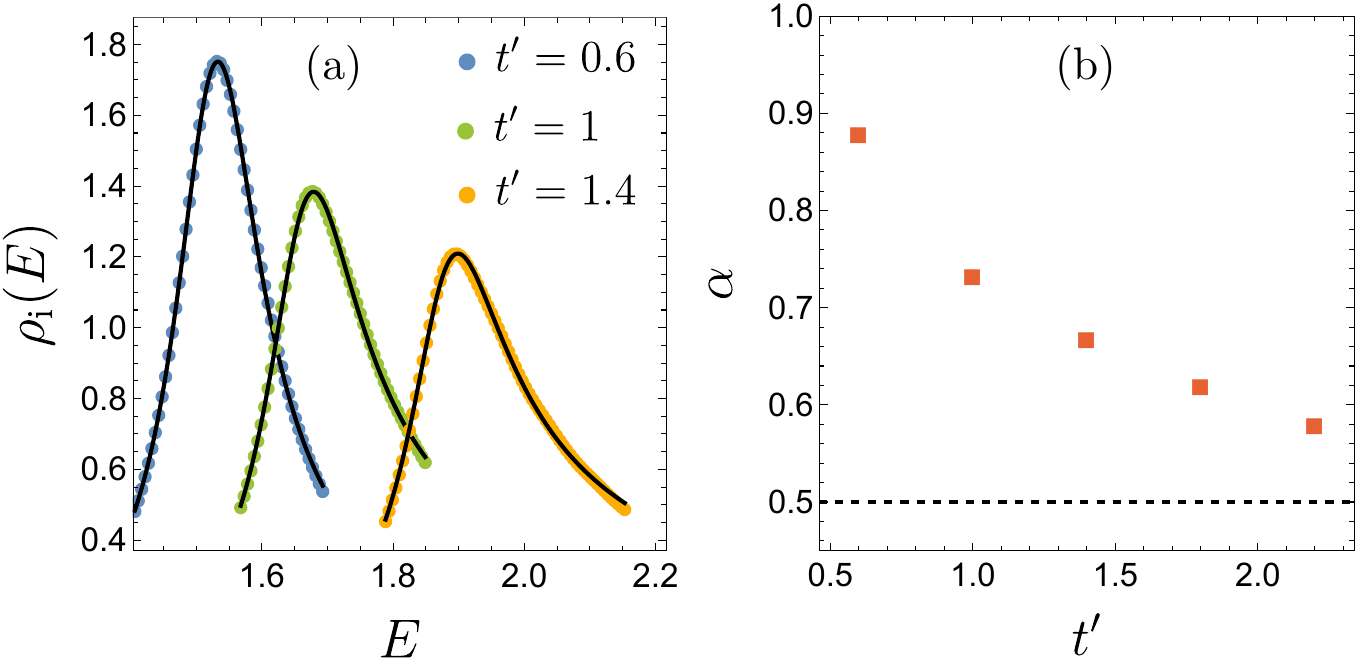}
\caption{Near-threshold behavior of the impurity TDOS for $\eta=0.075$. Other parameters are as in Fig.~\ref{fig3}, where differences can be traced back to the different values for
$\eta$. (a) Data points represent DMRG results for $\rho_{\rm i}(E)$ for several $t'$. Solid lines are fits to Eq.~(\ref{fitformula}) with $A$, $\Delta_d$, and $\alpha$ as fitting parameters. (b) Power-law exponent $\alpha$ vs $t'$ obtained from the fits. The dashed line shows $\alpha=1/2$ as predicted by field theory. } \label{fig4}
\end{figure}

As independent approach, we have performed DMRG simulations \cite{Hallberg2006} to compute the TDOS for the half-filled Hubbard model in Fig.~\ref{fig1}(a), see Eq.~\eqref{model}.  
The ground state was obtained by finite-system DMRG in the matrix product state (MPS) formulation \cite{Schollwoeck2011}.
Local spectral functions were computed in energy space via the correction vector (CV) method \cite{Kuehner1999}. For details, see the Appendix.

Setting $t_0=1$, we focus on a Hubbard chain of length $L=23$ with $U=10$, for which Eq.~\eqref{gap} gives $\Delta=3.27$, and fix $U_d=3$. To obtain smooth curves, we replace $\delta$-function peaks in the finite-chain TDOS by Lorentzians with broadening $\eta\sim 1/L$. Typical numerical results for the TDOS at sites $j=0,1,2$ are shown in Fig.~\ref{fig3}. Clearly, the subgap feature is most pronounced in the impurity TDOS, but it also appears at the chain sites with a spectral weight that decays with distance from the boundary.  Moreover, the TDOS is consistent with a linear energy dependence for $E\agt\Delta$ as predicted by field theory.  

To quantitatively test the spinon-Kondo effect, we analyze the subgap peak in $\rho_{\rm i}(E)$,  taking into account broadening in the numerical data. 
A convolution of the power law in Eq.~(\ref{subgapregime}) with a Lorentzian yields the fitting function (see the Appendix)
\be\label{fitformula}
f(E)=A\text{ Im}[(\Delta_d-E-i\eta)^{-\alpha}], 
\ee
where $A$ is a nonuniversal prefactor. To extract $\alpha$, we should in principle fit the energy dependence for $|E-\Delta_d|\ll T_K$. However, for small $t'$, we may have $T_K\ll \eta$, such that the broadening obscures the threshold singularity. Equivalently, since $\eta\sim 1/L$,  for $t'\ll t_0$, the chain length is  smaller than the size of the Kondo screening cloud \cite{Laflorencie2008} and the RG flow to the strong-coupling point is cut off by finite-size effects.  We then 
expect an effective phase shift $\gamma<\pi/2$, resulting in a larger power-law exponent than the predicted value $\alpha=1/2$, see Eqs.~(\ref{subgapregime}) and (\ref{plaw}). This consideration should be accounted for in STS experiments on relatively short chains \cite{Jacob2021,Zhao2025,Su2025}. 
For larger values of $t'$, however, $T_K$ increases and one thus expects the Kondo screening cloud to shrink. We therefore expect that $\alpha$ approaches the predicted value
$\alpha=1/2$ as one increases $t'$.

As shown in Fig.~\ref{fig4}(a), we find excellent agreement between DMRG data and Eq.~(\ref{fitformula}). From the fits, we extract the values of $\Delta_d$ and $\alpha$. First, we observe energy shifts $\Delta_d>U_d/2$ for $t'/t_0\geq 0.6$, where the line shape becomes more asymmetric with increasing $t'$.  Second, the $t'$-dependence of $\alpha$  is shown in Fig.~\ref{fig4}(b). Indeed, the numerical estimates for $\alpha$ approach $\alpha=1/2$ with increasing $t'$, where 
the subgap feature in $\rho_{\rm i}(E)$ is governed by the spinon-Kondo effect at strong coupling.

\section{Discussion and conclusions}\label{sec5}

The spinon-Kondo effect gives rise to a universal power-law singularity in the subgap TDOS of a 1D Mott insulator in the presence of a magnetic impurity, different from a conventional Lorentzian Kondo resonance of width $T_K$ around zero energy. We have focused on the regime $\Delta_d<\Delta$, where the predicted 
asymmetric singularity, see Fig.~\ref{fig1}(b), should be readily observable by STS for 1D nanographene \cite{Mishra2021,Zhao2024a,Zhao2025} or molecular \cite{Sun2025,Su2025} chains. 
Such observations can yield direct evidence for the spinon-Kondo effect.
For $\Delta_d>\Delta$, the singular features overlap with the bulk continuum spectrum and may be harder to observe.  

\new{We note that for the corresponding case of a band insulator, one in addition has a bulk spin gap $\Delta_s$.  As a consequence, the Kondo effect is absent because the bulk modes relevant for establishing Kondo screening processes are fully gapped out \cite{Hewson1993}. In the limit $T_K\ll \Delta_s$ (with $T_K$ defined for $\Delta_s\to 0$), only a sharp TDOS peak at $|E|=\Delta_d$ is then found from the corresponding gapped Anderson impurity model, see Refs.~\cite{Chen1998,Moca2010} for a detailed discussion.}

Finally, let us comment on the 2D case. In Ref.~\cite{He2022}, a subgap peak was found within a parton mean-field approach for the spinon spectral function of a  U(1) quantum spin liquid in 2D. However, Ref.~\cite{He2022} studied the regime $\Delta_d>\Delta$ by invoking additional holon-spinon interactions. 
Beyond parton mean-field theory, 2D spinon Fermi surface states are strongly coupled non-Fermi liquid states \cite{Ribeiro2011,Lee2009,Metlitski2010}. We speculate (see the Appendix for details) that TDOS singularities for $\Delta_d<\Delta$ could also be used to probe the anomalous scaling dimension of spinons in 2D.

\emph{Data availability.---}All data underlying the figures in this paper are available at Zenodo \cite{Zenodo}.

\begin{acknowledgments} 
We thank N. Andrei and R. Fasel for discussions. We
 acknowledge funding by the Deutsche Forschungsgemeinschaft (DFG, German Research Foundation) under Projektnummer 277101999 - TRR 183 (project C01) and under Germany's Excellence Strategy - Cluster of Excellence Matter and Light for Quantum Computing (ML4Q) EXC 2004/1 - 390534769. This work was supported by a grant from the Simons Foundation (Grant No. 1023171, R.G.P.), by Finep
(Grant No. 1699/24 IIF-FINEP, R.G.P.), and Conselho Nacional de Desenvolvimento Cient\'{i}fico e Tecnol\'{o}gico – CNPq (R.G.P.).   R.E. acknowledges support from the ``Maldacena Program for Visiting Professors'' at the Balseiro Institute.
\end{acknowledgments}

\appendix

\section{Effective Hamiltonian}

In this Appendix, we provide technical details on the effective model. Using the model \eqref{model}, let us derive the effective Hamiltonian in the charge sector with one additional electron with respect to the half-filled system using a canonical transformation in the regime $t'\sim t_0 \ll U_d\lesssim U$. The idea is to eliminate all terms that create double occupancies at first order in $t_0,t'$  and subsequently project onto the charge states at the specified filling. The transformed Hamiltonian has the form $\tilde H=V+H_{\rm spin}+T_0+\mc{O}(t_{0}^3/U^2)$. Here, $V$  comprises the interaction terms of Eq.~\eqref{model} and $H_{\rm spin}$ is the Heisenberg exchange interaction in Eq.~(\ref{spinchain}) with $\mb S_j=\frac12\sum_{\sigma,\sigma'}c^\dagger_{j\sigma} \boldsymbol{\sigma}_{\sigma\sigma'} c^{\phantom\dagger}_{j\sigma'}$. The constrained hopping term is given by 
\bea
T_0&=&-\sum_{\sigma}\sum_{j=0}^{L-1} \kappa_j (n_{j\bar \sigma}c^\dagger_{j\sigma}c^{\phantom\dagger}_{j+1,\sigma} n_{j+1,\bar \sigma}\nonumber\\
&&+\bar n_{j\bar \sigma}c^\dagger_{j\sigma}c^{\phantom\dagger}_{j+1,\sigma}\bar n_{j+1,\bar \sigma}+\text{h.c.}),
\eea
where $\bar \sigma=-\sigma$, $\bar n_{j\sigma}=1-n_{j\sigma}$, and the position-dependent hopping parameter is $\kappa_0=t'$ and  $\kappa_j=t_0$ for $j\geq 1$. 
Under the  canonical transformation, the electron annihilation operator becomes 
\bea
&&\tilde c_{j\sigma}=X_{j\sigma}c_{j\sigma}+\frac{2\kappa_j}{U_j+U_{j+1}}\left(n_{j\bar\sigma}-n_{j+1,\bar\sigma}\right)c_{j+1,\sigma}\nonumber\\
&&+\frac{2\kappa_{j-1}}{U_j+U_{j-1}}\left(n_{j\bar\sigma}-n_{j-1,\bar\sigma}\right)c_{j-1,\sigma}+\mc O(\kappa_j^2/U_j^2),
\eea
where  $U_0=U_d$ and $U_j=U$ for $j\geq 1$, with 
\bea
X_{j\sigma}&=&1+\frac{2\kappa_j}{U_j+U_{j+1}}\left(c^\dagger_{j\bar\sigma}c^{\phantom\dagger}_{j+1,\bar\sigma}-c^\dagger_{j+1,\bar\sigma}c^{\phantom\dagger}_{j\bar\sigma}\right)\nonumber\\
&&+\frac{2\kappa_{j-1}}{U_j+U_{j-1}}\left(c^\dagger_{j\bar\sigma}c^{\phantom\dagger}_{j-1,\bar\sigma}-c^\dagger_{j-1,\bar\sigma}c^{\phantom\dagger}_{j\bar\sigma}\right).
\eea 
The projection of $T_0+V$ onto the subspace with only one doubly occupied site yields an effective tight-binding model for the charge sector,
\bea
H_{\rm charge}&=&\frac{U_d}2 d_0^\dagger d^{\phantom\dagger}_{0}-t'(d^\dagger_1d^{\phantom\dagger}_{0}+\text{h.c.})\nonumber\\
&&+\sum_{j\geq 1}\left[\frac{U}2 d^\dagger_jd^{\phantom\dagger}_{j}-t_0(d^\dagger_jd^{\phantom\dagger}_{j+1}+\text{h.c.})\right],
\eea 
where $d_j^\dagger$ is the fermionic creation operator for a spinless holon at site $j$. 
As a result, for $U_d<U$, the holon is attracted to the impurity site, which leads to a holon bound state. For $t'\ll U-U_d$, second-order perturbation theory yields 
the bound state energy $U/2+\delta E_c$ with  
\be
\delta E_c\approx -4(t')^2/(U-U_d)<0.
\ee

\section{Bulk magnetic impurity}

We here briefly describe the corresponding case of a magnetic impurity embedded in the bulk of the 1D Mott insulator.  Assuming a symmetric coupling of the impurity to both sides of the chain, one arrives at a two-channel Kondo scenario, where the Kondo coupling flows to an overscreened non-Fermi liquid fixed point \cite{Affleck1991}.  (Since the two-channel Kondo fixed point is unstable against channel asymmetry, the single-channel spinon-Kondo effect discussed in the main text is recovered for asymmetric couplings.)  In the presence of a holon bound state corresponding to the fermion creation operator $d_{\rm bs}^\dagger$, the boundary electron is now fractionalized as \cite{Affleck1995}
\be
c^\dagger_{0\uparrow}\sim d_{\rm bs}^\dagger \phi^{(\frac12)}(0) \sigma(0) , 
\ee
where conformal field theory (CFT) \cite{Affleck1991,Affleck1993} yields the chiral spinor $\phi^{(\frac12)}(x)$ of the SU(2)$_2$ Wess-Zumino-Witten model \cite{Gogolin1998} with conformal dimension $3/16$.
The order parameter $\sigma(x)$ of the Ising CFT has conformal dimension $1/16$.  Computing the TDOS as in the main text, we then again arrive at Eq.~\eqref{subgapregime}.  
Remarkably, the universal power-law exponent $\alpha=1/2$ of the subgap TDOS singularity is thus expected both for boundary and embedded impurities. 

\section{Details on DMRG simulations}

Next, let us provide details about our DMRG simulations \cite{Hallberg2006} for the model  \eqref{model}. 
The ground state $|\Omega\rangle$ and energy $E_0$ were obtained with finite-system DMRG in the MPS formulation (5 sweeps, $m=1024$, energy error $\sim10^{-11}$) \cite{Schollwoeck2011}, and local spectral functions were computed in energy space via the CV method \cite{Kuehner1999}. For the greater Green's function,
\be
G^{>}_{j \sigma}(E)=\langle\Omega|c_{j\sigma}\frac{1}{E+i\eta+E_0-H} c^\dagger_{j\sigma}|\Omega\rangle,
\ee
we solve the linear problem 
\be
(E+i\eta+E_0-H)|X \rangle=|a\rangle,\quad  |a\rangle= c^\dagger_{j\sigma}|\Omega\rangle,
\ee
such that $G^{>}_{j\sigma}(E)=\langle a|X\rangle$. Since the system has particle-hole and spin symmetry, the TDOS follows as $\rho(j,E)=-\tfrac{2}{\pi}{\rm Im}\, G^{>}_{j\sigma}(E)$ with arbitrary   $\sigma$.
Numerically, we solve the equivalent real positive-definite system 
\be 
[(E+E_0-H)^2+\eta^2]|X_I\rangle=-\eta |a\rangle,
\ee
and reconstruct $|X\rangle=|X_R\rangle+i|X_L\rangle$ through 
\be 
|X_R\rangle=-\eta^{-1}[H-E_0-E]|X_I\rangle.
\ee

Our iterative solution uses the restarted generalized minimal residual method (GMRES) with tolerance $\sim 10^{-4}$ and inner restart $\sim 512$.
The component $|X_I\rangle$ from the previous energy (or position) is used as initial guess to accelerate convergence. 
Within DMRG, each sweep position constructs the local effective operator, solves for the CVs (real and imaginary parts), and includes them as targets with equal weights in the reduced density matrix used for truncation. The resulting eigenvectors then define the renormalized basis for the next step. Using small values for the broadening parameter $\eta$ increases resolution but degrades GMRES conditioning and requires stricter solver tolerances.  For the reported values of $\eta$, the solver tolerance is $\sim 10^{-4}$.
For dynamical calculations, we used 5 sweeps and $m=256$, which results in an error $\sim 10^{-8}$ in the Green's functions and the TDOS. 

\begin{figure}[t]
        \includegraphics[width=.9\columnwidth]{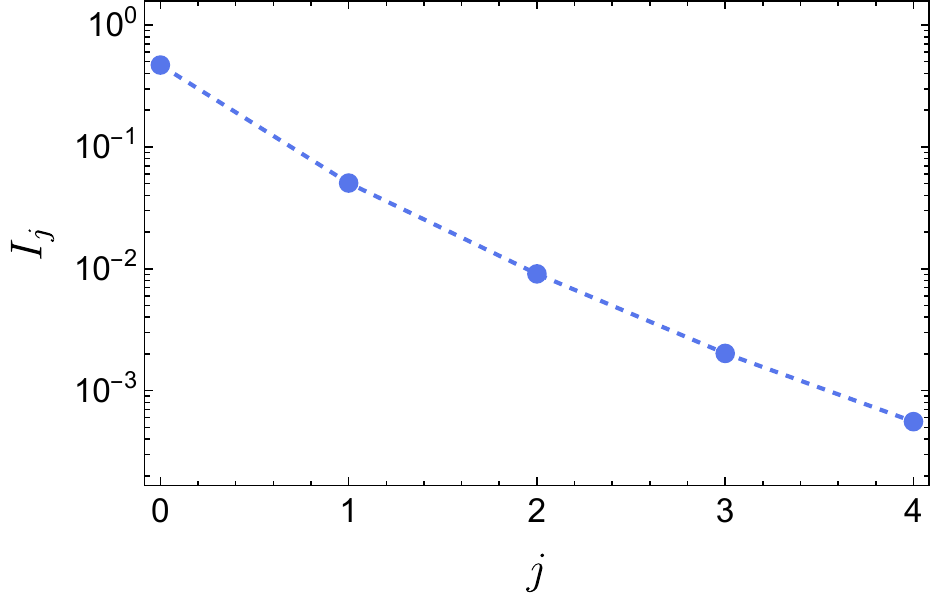}
    \caption{Integrated spectral weight $I_j$ of the subgap feature vs site index $j$ obtained from DMRG simulations for $L=23$ with $t_0=1$,  $t'=0.6$, $U=10$, $U_d=3$, and  $\eta=0.025$. Note the semi-logarithmic axes.  The dashed curve is a guide to the eye only. } \label{fig5}
\end{figure} 

In addition to the numerical results shown in the main text, we have used DMRG simulations to calculate the TDOS for other choices of the parameters $U$, $U_d$, $t'$, and $\eta$. As a rule, we found that the subgap features are present for small values of $U_d/U$. As one increases $U_d/U$, this asymmetric subgap peak starts to significantly overlap with the continuum above the bulk gap $\Delta$. We also observe that the sign of the energy shift $\Delta_d-U_d/2=\delta E_s+\delta E_c$ depends on the detailed values of the parameters. While in Figs.~\ref{fig3} and \ref{fig4}, we have focused on the regime $t'\geq 0.6t_0$ where $\Delta_d>U_d/2$,  we find  $\Delta_d<U_d/2$ for smaller values of $t'/t_0$. 

We next note that in order to analyze numerical results for the TDOS threshold behavior at energies $|E|\agt \Delta_d$, we describe the broadening of the subgap peak using a convolution of the power law \eqref{subgapregime} with a  Lorentzian of width $\eta$,
\be
f(E)=A\int_{-\infty}^{\infty}\frac{dz}{\pi}\,  \frac{\eta}{(E-z-\Delta_d)^2+\eta^2} \frac{\Theta(z)}{z^\alpha},
\ee
which leads to   Eq.~(\ref{fitformula}).
Let us also note that the TDOS obeys the sum rule $\int_0^{\infty}dE\, \rho_\sigma(j,E) =\frac12$ at half filling. In the regime where the subgap feature is well separated from bulk contributions, we measure the weight of the subgap peak as function of the distance from the impurity by computing the integral $I_j=\int_0^\Delta dE\,\rho_\sigma(j,E)$. As shown in Fig.~\ref{fig5}, our numerical results indicate that $I_j$ decays exponentially with $j$, confirming that the subgap feature is associated with a boundary bound state.   

In addition, we observe that for $j\geq 2$, the TDOS exhibits oscillations as a function of energy. In particular, for $j=2$ in Fig.~\ref{fig3}, the TDOS is suppressed at one minimum with energy $E>\Delta$, but we  find that the number of minima of $\rho_\sigma(j,E)$ increases with $j$. To interpret this result, we have verified that similar oscillations appear in the noninteracting model, $U=U_d=0$, where they can be traced back to the $\sim\sin(kja)$ dependence of the electron wave function near the boundary, with the momentum $k$ being governed by the energy of the electron injected into the chain. In the interacting case, we expect a similar behavior due to the wave function of holon scattering states near the boundary. More precisely, the TDOS involves nontrivial matrix elements between the exact eigenstates of the Hubbard model, but  analytical expressions are not available for analyzing the resulting energy dependence in detail.\\ 

\section{Toward the 2D case}

Finally, moving toward generalizations of our approach to 2D Mott insulators, 
let us briefly comment on the  slave rotor mean-field approach to the spinon-Kondo effect \cite{Florens2002,Florens2004,He2022}.  Here one adopts the fractionalization scheme $c_{j\sigma}\sim f_{j\sigma} e^{-i\theta_j}$, where $f_{j\sigma}$ is a spinful fermionic annihilation operator (spinon) and $\theta_j$
is a bosonic field associated with charge degrees of freedom (chargon).  The physical electron operator $c_{j\sigma}$ is invariant under U(1) gauge transformations, $f_{j\sigma}\mapsto
e^{i\alpha_j}f_{j\sigma}$ and $\theta_j\mapsto\theta_j+\alpha_j$, with $\alpha_j\in\mathbb{R}$.
The charge fluctuation operator $L_j=\sum_\sigma c_{j\sigma}^\dagger c_{j\sigma}^{}-1$ plays the role of an angular momentum operator canonically conjugate to $\theta_j$ since 
$[\theta_j,L_{j'}]=i\delta_{jj'}.$ 
At the mean-field level, all constraints involving charge and spin operators are implemented on average by Lagrange multipliers, and one neglects interactions mediated by fluctuations of the U(1) gauge field. The mean-field parameters and the corresponding chargon and spinon dispersions are  determined by self-consistency equations. 

In the Mott insulator phase, chargons are gapped and spinons form a Fermi surface.  The TDOS can then be calculated by a convolution of spinon and chargon Green's functions.    In Ref.~\cite{He2022}, a subpeak peak with no intrinsic width   was obtained by  including  a sufficiently strong short-range spinon-chargon interaction, treated within the random phase approximation.  Beyond this approximation, we expect the spinon Green's function  to be affected by interactions with the U(1) gauge field. In fact, the anomalous dimension of a fermionic spinon has been calculated in a large-$N$ expansion \cite{Metlitski2010}. Such an anomalous dimension should affect the energy scaling, potentially turning the subgap peak in the 2D quantum spin liquid into a threshold singularity analogous to the 1D case. Our results for the 1D case also suggest considering magnetic impurities in the regime $\Delta_d<\Delta$, in which case the  presence of the subgap peak does not rely on the assumption of strong spin-charge interactions.

\bibliography{biblio}
\end{document}